\documentclass[conference, 10pt]{IEEEtran}

\usepackage{spconf}

\usepackage{graphicx}
\usepackage{amsmath}
\usepackage{mathtools}

\usepackage{amsmath}

\usepackage{bbm}
\newcommand{\vect}[1]{\boldsymbol{#1}}

\usepackage{breqn}

\newcommand\thefontsize{The current font size is: \f@size pt}

\IEEEoverridecommandlockouts
\usepackage{cite}
\usepackage{amsmath,amssymb,amsfonts}
\usepackage{graphicx}
\usepackage{textcomp}
\usepackage{xcolor}
\def\BibTeX{{\rm B\kern-.05em{\sc i\kern-.025em b}\kern-.08em
    T\kern-.1667em\lower.7ex\hbox{E}\kern-.125emX}}

\usepackage[font={small}]{caption, subfig}

\usepackage{algorithm}
\usepackage[noend]{algpseudocode}

\usepackage{footnote}

\newcommand\blfootnote[1]{%
  \begingroup
  \renewcommand\thefootnote{}\footnote{#1}%
  \addtocounter{footnote}{-1}%
  \endgroup
}

\algrenewcommand\algorithmiccomment[1]{\hfill \(\phantom{ }\) #1}

\newcommand{\algorithmfootnote}[2][\footnotesize]{%
  \let\old@algocf@finish\@algocf@finish
  \def\@algocf@finish{\old@algocf@finish
    \leavevmode\rlap{\begin{minipage}{\linewidth}
    #1#2
    \end{minipage}}%
  }%
}

\begin{document}

\title{Earthquake Location and Magnitude Estimation \\ with Graph Neural Networks}

\maketitle
\thispagestyle{plain}
\pagestyle{plain}

\begin{abstract}

We solve the traditional problems of earthquake location and magnitude estimation through a supervised learning approach, where we train a Graph Neural Network to predict estimates directly from input pick data, and each input allows a distinct seismic network with variable number of stations and positions. We train the model using synthetic simulations from assumed travel-time and amplitude-distance attenuation models. The architecture uses one graph to represent the station set, and another to represent the model space. The input includes theoretical predictions of data, given model parameters, and the adjacency matrices of the graphs defined link spatially local elements. As we show, graph convolutions on this combined representation are highly effective at inference, data fusion, and outlier suppression. We compare our results with traditional methods and observe favorable performance.

\end{abstract}

\vspace{0.1 cm}

\section{Introduction}

Earthquake location and magnitude estimation are two longstanding challenges in seismology. These parameters are estimated for millions of earthquakes every year using data from many regional monitoring networks around the world. Solutions are usually framed in well understood signal processing and Bayesian inverse theory contexts \cite{tarantola1982generalized}. By their nature, `traditional' inverse methods have certain subjective biases, such as assuming fixed parameter likelihood functions (e.g., Gaussian kernels), stacking estimates over stations, or assuming uncorrelated noise between stations (i.e., neglecting epistemic uncertainties due to physical model errors and local correlated sources of noise). These weaknesses induce errors and biases in catalogs resulting point estimates of source parameters, that are generally weakly correlated in space and time \cite{kagan2003accuracy}. 

\blfootnote{\tiny{Copyright disclaimer: © 2022 IEEE. Personal use of this material is permitted. Permission from IEEE must be obtained for all other uses, in any current or future media, including reprinting/republishing this material for advertising or promotional purposes, creating new collective works, for resale or redistribution to servers or lists, or reuse of any copyrighted component of this work in other works. Conference proceedings limit manuscript length to 4 pages. Correspondence address: imcbrear@stanford.edu.}}.

When applied to the same type of traditional inverse problem, Graph Neural Networks (GNN), and graph signal processing in general \cite{ortega2018graph}, offer a powerful new approach to arrive at stable global model estimates of parameters (i.e., location and magnitude) through processing complex, noisy data, on irregularly distributed station networks. As a graph is defined as a set of nodes, with an \textit{adjacency matrix} that specifies which pairs of nodes have a (directed) edge linking them - if the nodes represent data points - the adjacency matrix gives access to an object that behaves in a similar way as a covariance matrix (with off diagonal terms included) encountered in traditional signal processing and Bayesian inverse theory. Since the adjacency matrix links a subset of nearby nodes together, and algorithms are applied to `collect and aggregate' information between local neighborhoods of the graph, these tools allow utilizing local coherency and redundancy between nodes to, for example: denoise and stabilize measurement estimates, hierarchically cluster and weight data to obtain global estimates of model parameters, or suppress the effect of data outliers \cite{djuric2018cooperative}.

We use Graph Neural Networks to solve the traditional point estimation problems of location and magnitude from arrival time and amplitude data. This approach combines the strengths of graphs and the capability of deep learning to predict the target quantities directly from input data, rather than relying on globally optimizing an explicitly posed objective function \cite{bergen2019machine, mousavi2020bayesian, van2020automated, munchmeyer2021earthquake}. Advantages of training a `surrogate model' to predict solutions from data, are many, such as (i) the estimates are available immediately without need to apply an optimizer, (ii) the estimates can be more robust in the presence of high noise and outlier data points, and (iii) the model can be trained to account for variable seismic network geometry. 

Finally, as a major component of our architecture, we introduce an important type of graph: the \textit{Cartesian product} graph of $Stations \times Model \text{ } Space$. This graph represents data measured over all stations, and over all points of model space, where the model space is the $Source \text{ } Region$ or $Magnitude \text{ } Axis$. It effectively represents information about all aspects of the problem (observed data, and predicted data, given model parameters). By giving these graphs structure, through local edge schemes that connect nearby points in model space and nearby stations, we allow local interaction between both model space points and station data during iterative graph convolutions. As we show, the pairwise structure $Stations \times Model \text{ }Space$ is the natural object on which observed data can be parsed, and thus convolving directly on this structure should improve inference.

\section{Methods}

\subsection{Graph Message Passing}

The basic computational unit of a standard GNN is the generic \textit{message passing} layer. The message passing layer captures the powerful notion of \textit{graph convolution} (e.g., \cite{hamilton2017inductive, battaglia2018relational}), wherein the latent states of nodes on graphs are updated based on (learned) mappings from the latent states of neighboring nodes. The operator is generically defined as $\mathcal{G}(\vect{h}_{\mathcal{A}}, \mathcal{E}_{\mathcal{A}' \leftarrow \mathcal{A}}) : \mathcal{A} \longrightarrow \mathcal{A}'$, and is given as

\vspace{-0.35 cm}

\begin{equation} \label{eq: forward map}
\begin{split}
    \mathcal{G}(\vect{h}_{\mathcal{A}}&, \mathcal{E}_{\mathcal{A}' \leftarrow \mathcal{A}}) = \\ 
    & \phi^{agg} \bigl( \vect{h}_{i}, \text{POOL} \{\phi^{msg}(\vect{h}_j, \vect{e}_{ij}, \vect{z}) \mid j \in \mathcal{N}(i) \} \bigr),
\end{split}
\end{equation}

\noindent which maps a latent signal $\vect{h}_{\mathcal{A}} \in \mathbb{R}^{|\mathcal{A}|\times K}$ measured on graph $\mathcal{A}$ (of feature dimension $K$) to a latent signal $\vect{h}_{\mathcal{A}'} \in \mathbb{R}^{|\mathcal{A}'| \times K'}$ measured on graph $\mathcal{A}'$ (of feature dimension $K'$), with directed edges $\mathcal{E}_{\mathcal{A}' \leftarrow \mathcal{A}}$ pointing nodes from $\mathcal{A}$ into $\mathcal{A}'$. When $\mathcal{A}' \neq \mathcal{A}$, this is a bipartite mapping between two distinct graphs; when $\mathcal{A}' = \mathcal{A}$, it is regular graph convolution within a single graph ($\mathcal{E}_{\mathcal{A}} \equiv \mathcal{E}_{\mathcal{A}' \leftarrow \mathcal{A}}$ when $\mathcal{A}' = \mathcal{A}$), yet either case is easily represented by this operator.

In the forward call \eqref{eq: forward map}, on a per-node basis $(i)$, the latent features of each neighboring node $j \in \mathcal{N}(i)$ are collected, and transformed by the learnable Fully Connected Network (FCN) layer, $\phi^{msg}$, prior to being \textit{pooled} over the entire neighborhood, $\mathcal{N}(i)$, where the pooling is over the node axis ($POOL : \mathbb{R}^{|\mathcal{N}(i)| \times K} \longrightarrow \mathbb{R}^{K}$), and is any global pooling operator such as \textit{sum-, mean, or max-pool}. The pooled aggregate message is then concatenated with the self state of node $\vect{h}_{i}$, and transformed by an additional learnable FCN, $\phi^{agg}$, resulting in the updated state on node $i$. Inside the message passing layer $\phi^{msg}$, the optional term of $\vect{e}_{ij}$ represents edge data between two nodes $(i,j)$, such as Cartesian offsets of node coordinates (if nodes represent spatially located points), and $\vect{z} = \text{POOL}\{\phi^{glb}(\vect{h}_j)\}$ is an optional `global' summary feature, that can be extracted from the entire graph at each step, using an additional learnable FCN, $\phi^{glb}$. By combining these facets, the overall capability of the graph convolution cell is to learn local feature transformations that work well for a wide range of nodes, while also exploiting local information transfer through the adjacency matrix. 

\subsection{Adjacency Matrices}

In our method there is one component (or graph) that specifies the station set, $\mathcal{S}$, and another that represents the model space, $\mathcal{H}$, where we consider either source region $\mathcal{H} = \mathcal{X}$ or magnitude axis $\mathcal{H} = \mathcal{M}$. For either model space, we simply need a (regular or irregular) grid that spans the region of interest. For $\mathcal{X}$, we take a collection of $\sim$100's of sparsely distributed points covering the source region of interest, and for the magnitude axis we take a fine spacing of nodes in magnitude (e.g., [-3,7] M with 0.1 or 0.25 M increment). 

Rather than leave the station set and model spaces as unordered sets, we give these sets structure by interpreting them as graphs. Specifically, $(\mathcal{S}, \mathcal{E}_{\mathcal{S}})$ is the station graph, $(\mathcal{X}, \mathcal{E}_{\mathcal{X}})$ is the source graph, and $(\mathcal{M}, \mathcal{E}_{\mathcal{M}})$ is the magnitude graph, where we define \textit{edge sets}, $\mathcal{E}_{\mathcal{S}}, \mathcal{E}_{\mathcal{X}}, \mathcal{E}_{\mathcal{M}}$, or equivalently adjacency matrices, that specify which elements in the respective sets are linked. While many edge construction schemes are possible \cite{berton2018comparison}, GNN's often work well for a wide range of adjacency matrix choices \cite{zhou2020graph} because the GNN can learn how to use whichever distribution of graphs it is trained over. For point cloud data, such as represented in $\mathcal{S}$ and $\mathcal{X}$, a natural choice of \textit{edge construction scheme} is to use K-nearest-neighbor (K-NN) graphs or $\epsilon-$distance graphs \cite{berton2018comparison}, since these favor connecting nearby elements to one another. For $\mathcal{M}$, while it is simply a linear grid of nodes, $K-NN$ and $\epsilon-$distance graphs effectively represent 1D convolution kernels over the $\mathcal{M}$ axis, and are hence also a natural choice. 

For our purposes we use $K-NN$ instead of $\epsilon-$distance graphs for all of $\mathcal{S}$, $\mathcal{X}$, $\mathcal{M}$, since these limit the incoming degree of all nodes to $K$, whereas $\epsilon-$distance graphs can have substantially different degrees across different nodes of the graph. While we did not optimize these choices fully, modest trial and error led to

\begin{subequations}
\begin{equation}
    \mathcal{E}_{\mathcal{S}} = \{(i,j) \mid \vect{s}_j \in K\text{-}NN_{8}(\vect{s}_i)\}
\end{equation}
\begin{equation}
    \mathcal{E}_{\mathcal{X}} = \{(i,j) \mid \vect{x}_j \in K\text{-}NN_{15}(\vect{x}_i)\}
\end{equation}
\begin{equation}
    \mathcal{E}_{\mathcal{M}} = \{(i,j) \mid \vect{m}_j \in K\text{-}NN_{10}(\vect{m}_i)\}
\end{equation}
\end{subequations}

\noindent where $K-NN(\cdot)_K$ represent the $K$-nearest-neighbors for the input node. Note that, these `edge sets' specify which entries in all possible $N \times N$ pairs for a graph of $N$ nodes are linked, and that the relationship is generally not symmetric (i.e., $(i,j) \in \mathcal{E}$ does not necessarily imply $(j,i) \in \mathcal{E}$). Edge sets are hence equivalent to adjacency matrices, but contain the same data in a more compact form.

We still need to define one essential graph type: the \textit{Cartesian product} graph of $\mathcal{S} \times \mathcal{H}$, which has as nodes all pairs of station-model ($\vect{s}$, $\vect{h}$) space coordinates. As discussed earlier, and as highlighted in the next section, it is most naturally all pairs of $(\vect{s}, \vect{h}) \in \mathcal{S} \times \mathcal{H}$ stations, and model space coordinates, that we can use to parse the data. The graphs already defined can be used directly to construct the Cartesian product graph with two-edge types, given by

\begin{subequations} \label{eq: cartesian product graph}
\begin{equation}
    \mathcal{E}_{\mathcal{H} \leftarrow \mathcal{H}, \mathcal{S}} = \bigl\{ (i,j) \mid \vect{h}_j \in \mathcal{N}(\vect{h}_i) \wedge (\vect{s}_j = \vect{s}_i) \bigr\}
\end{equation}
\begin{equation}
    \mathcal{E}_{\mathcal{S} \leftarrow \mathcal{S}, \mathcal{H}} = \bigl\{ (i,j) \mid \vect{s}_j \in \mathcal{N}(\vect{s}_i) \wedge (\vect{h}_j = \vect{h}_i) \bigr\}.
\end{equation}
\end{subequations}

\noindent As can be seen in \eqref{eq: cartesian product graph}, these edge types capture two different types of relationships between nodes on the Cartesian product: connecting either (i) a fixed station, and two neighboring source nodes (or magnitude nodes), or (ii) a fixed source node (or fixed magnitude node), and two neighboring station nodes. Intuitively, the alternative relationships offer complementary insights into the local and global features between different nodes on the graph. While the cardinality of $\mathcal{S} \times \mathcal{H}$ is high (e.g., 250 stations $\times$ 200 spatial nodes $=$ 50,000), the incoming degree of each node (for either edge type) is limited by the choices of $K$ for any of the component graphs, $\mathcal{S}, \mathcal{X}, \mathcal{M}$, and hence these graphs are highly sparse.

\subsection{Input Features}

The input features we pass to the GNN are defined for arbitrary input pick data sets, where for location, it includes pick times (with assigned phase types), and for magnitude it includes amplitude measurements (with assigned phase types). Each input can be any given station set, $\mathcal{S}$, and individual stations in this set can still have missing picks or false picks. For the source location problem our input tensor is defined analogously to the `pre-stack' back-projection metric (e.g., \cite{ringdal1989multi}), which, for each spatial coordinate $\vect{x} \in \mathcal{X}$ is the measure of how close the nearest arrival, $\tau_i^k \in \mathcal{D}_i$, of phase type $k$ on station $\vect{s}_i \in \mathcal{S}$, is to the theoretical arrival time from that source coordinate. Hence, we compute

\begin{equation} \label{eq: input tensor location}
    \vect{h}_{k}^{\mathcal{X}}(\vect{s}_i, \vect{x}) = \exp\Biggl(-\frac{\bigl(t_0 + T_k(\vect{s}_i, \vect{x}) - \tau_i^k\bigr)^2}{2\sigma_t^2}\Biggr) 
\end{equation}

\noindent where $T_k(\vect{s}_i,\vect{x})$ is the theoretical travel time calculator, for phase types $k$, and $\sigma_t$ is a fixed kernel wide enough to capture a range of misfit times appropriate to the scale of the application (e.g., $\sigma_t =$ 5 s). Travel times are computed with the Fast Marching Method \cite{sethian1996fast} using the regional 3D velocity model of \cite{thurber2009regional}.

For the magnitude input, a similar input scheme is defined. We compute

\begin{equation} \label{eq: input tensor magnitude}
    \vect{h}_{k}^{\mathcal{M}}(\vect{s}_i, \vect{m}) =  \exp\Biggl(-\frac{\bigl(A_k(\vect{s}_i, \vect{m}, \vect{x}) - \log_{10}(a_i^k) \bigr)^2}{2\sigma_a^2}\Biggr)
\end{equation}

\noindent where $A_k(\vect{s}_i, \vect{m}, \vect{x})$ predicts the theoretical $\log_{10}$ amplitude on station $\vect{s}_i$, from source $\vect{x} \in \mathcal{X}$, with magnitude $\vect{m} \in \mathcal{M}$, for phase type $k$. Hence, this feature measures for a given magnitude, how close the observed log-amplitude is to the theoretical log-amplitude. The Gaussian kernel width, $\sigma_a$, is application dependent, but is chosen to be large enough to capture a range of misfits in the feature space (e.g., $\sigma_{a}$ = 0.5). For our purposes we use a locally calibrated linear magnitude scale \cite{hutton1987ml}, that is fit to predict magnitudes consistent with the USGS in northern California, using maximum amplitude measurements around the times of P- and S-waves from the EHZ component of NC seismic network stations.

\subsection{Architecture}

The GNN architecture we propose is designed to be a simple, effective way to map the very high-dimensional input tensors $\vect{h}_{k}^{\mathcal{X}}(\vect{s}_i, \vect{x})$, $\vect{h}_{k}^{\mathcal{M}}(\vect{s}_i, \vect{m})$, which are feature vectors measured on all pairs of station-model space points, to explicit predictions of source likelihood $p(\vect{x}) \subset \Omega_{\mathcal{X}}$, and magnitude likelihood $p(\vect{m}) \subset \Omega_{\mathcal{M}}$, at any query points within the full continuous Euclidean model space, $\Omega_{\mathcal{H}}$. Each input can also be for a different number of stations, and a different station graph. 

To achieve this overall mapping, we first (i) apply repeated graph convolutions on the Cartesian product $\mathcal{S} \times \mathcal{H}$, then (ii) embed into the model space by stacking over stations (in the latent space, $\mathcal{S} \times \mathcal{H} \rightarrow \mathcal{H}$). Then, (iii) we repeat graph convolutions on $\mathcal{H}$, and finally (iv) we make \textit{predictions} of likelihood at arbitrary coordinates $p(\vect{h}_q) \subset \Omega_{\mathcal{H}}$, by using local masked-attention (i.e., effectively graph-based interpolation \cite{velivckovic2017graph}) of the queries $\vect{h}_q$ with respect to the fixed nodes in $\mathcal{H}$. This design is straightforward, and in a natural, physically motivated way, transforms the raw data measured over arbitrary station networks to predictions over arbitrary model space domains. The architecture is given in more detail in Algorithm (1).

\begin{algorithm} \label{alg:A_1}
\caption{The Architecture of the GNN}
\begin{algorithmic}[1] 
\Procedure{Input}{}
\State \textbf{Create graphs: } $\mathcal{S}$, $\mathcal{H}$, and $\mathcal{S} \times \mathcal{H}$.
\State \textbf{Measure features: } $\vect{h}^{\mathcal{H}}(\vect{s}_i, \vect{h})$
\State \textbf{Choose queries: } $\{\vect{h}_q \in \Omega_{\mathcal{H}} \}$.
\EndProcedure

\Procedure{Forward Pass \vspace{0.1 cm}}{}
\State \hspace{-0.4 cm} $\vect{h}_{\mathcal{S} \times \mathcal{H}}^{(l + 1)} = \phi \Bigl(\mathcal{G}_1\bigl(\vect{h}_{\mathcal{S} \times \mathcal{H}}^{(l)}, \mathcal{E}_{\mathcal{H} \leftarrow \mathcal{H}, \mathcal{S}}\bigr) , \mathcal{G}_2\bigl(\vect{h}_{\mathcal{S} \times \mathcal{H}}^{(l)}, \mathcal{E}_{\mathcal{S} \leftarrow \mathcal{S}, \mathcal{H}}\bigr) \Bigr)$
\Comment{\hfill [Repeat 3]}

\State \hspace{-0.4 cm} $\vect{h}_{\mathcal{H}}^{(l + 1)} = \mathcal{G}\bigl(\vect{h}_{\mathcal{S} \times \mathcal{H}}^{(l)}, \mathcal{E}_{\mathcal{H} \leftarrow \mathcal{S} \times \mathcal{H}}\bigr)$
\Comment{\hfill [Apply 1]}

\State \hspace{-0.4 cm} $\vect{h}_{\mathcal{H}}^{(l + 1)} = \mathcal{G}\bigl(\vect{h}_{\mathcal{H}}^{(l)}, \mathcal{E}_{\mathcal{H} \leftarrow \mathcal{H}}\bigr)$ \Comment{\hfill [Repeat 3]}
\EndProcedure

\Procedure{Prediction \vspace{0.1 cm}}{}

\State \hspace{-0.4 cm}  $\vect{h}_q^{pred} = \mathcal{G}\bigl(\vect{h}_{\mathcal{H}}^{(l)}, \mathcal{E}_{\vect{h}_q \leftarrow \mathcal{H}}\bigr)$ \Comment{\hfill [Apply 1]}

\EndProcedure

\end{algorithmic}

\small{Inside each $\mathcal{G}$, the learnable FCN's use feature dimensions between [15-30], and PReLU activation. The graph convolutions in steps [7,8,10] include edge data, $\vect{e}$, of Cartesian offsets between nodes, and in [8], a global state, $\vect{z} \in \mathbb{R}^5$ is also included. The edge set $\mathcal{E}_{\mathcal{H} \leftarrow \mathcal{S} \times \mathcal{H}}$ points nodes in the Cartesian product $\mathcal{S} \times \mathcal{H}$ into $\mathcal{H}$, by linking all stations for each fixed $\vect{h} \in \mathcal{H}$. $\mathcal{E}_{\vect{h}_q \leftarrow \mathcal{H}}$ is the $K-NN$ graph, which for each query $\vect{h}_q$ has edges linked to the $10$-nearest nodes in $\mathcal{H}$. The total number of free parameters of the 
GNN is $\sim$28,000.}

\end{algorithm}

\subsection{Training Details}

To train the models, we generate a diverse set of synthetic data, for the assumed physical travel time and local magnitude scale models. In either case, for a batch of 150 samples, for each sample, we (i) sample an arbitrary subset of 50 - 450 stations of the NC network, (ii) pick a random source coordinate uniformly over the domain (and random magnitude uniformly over the axis), (iii) compute theoretical arrival times and amplitude measurements, (iv) remove arrivals of source-receiver distances greater than $d + \epsilon$ with $d \sim$ Uniform [10, 1000] km per event, and $\epsilon \sim \mathcal{N}$(0, 30 km) per event and per station, (v) add per-pick Laplacian noise of std. 1$-$5$\%$ the quantity for travel time, and 25$-$250$\%$ the quantity for amplitude  (hence, uncertainties increase proportional to the observed quantity). Finally, we (vi) randomly delete 10$-$80$\%$ of arrivals, and randomly corrupt 10$-$30$\%$ of the remaining arrivals to have anomalous (uniformly random) arrival time and amplitude measurements. These simple steps can produce complex input datasets (including strong noise and anomalous picks), for variable seismic networks, with data that are still physically tied to the assumed travel time and magnitude scale models. 

For the labels, we represent each target solution as a continuous Gaussian, wherein the target spatial prediction is a multivariate Gaussian of fixed width $\sigma_{\mathcal{X}}$ (= 25 km), localized at the correct source coordinate, and similarly, the target magnitude prediction is a 1D Gaussian, localized on the true magnitude, with width $\sigma_{\mathcal{M}}$ (= 0.5 M). If a particular input sample has $\leq 3$ non-corrupted arrivals following the synthetic scheme above, the labels are set to all zero for that sample. Models are implemented in PyTorch Geometric \cite{fey2019fast}, and we train by computing L2-norm losses of the output with respect to the target over the batch, and optimizing with the Adam optimizer \cite{kingma2014adam} for $\sim$10,000 update steps with a learning rate of $10^{-3}$. In $\sim$10's of training runs, the GNN performed similarly and converged in training at similar rates, hence indicating that training was stable.

\section{Results}

To evaluate the performance of our method, we apply the trained GNN to locate and compute magnitudes of events for 1,500 new events generated through the synthetic data construction scheme described above. This allows measuring the bulk performance of the GNN as the available station network changes, source coordinates vary, and as the pick data has highly variable levels of noise, number of picks, and anomalous pick rates. We compare the results with traditional methods. For location, we minimize the unweighted least squares residual, using the average best solution of five particle swarm global optimization runs \cite{kennedy1995particle}. For magnitude, we compute magnitude estimates for all picks using our calibrated magnitude scale, and take the median of the set of points within the 10$-$90$\%$ quantile range of the distribution.

\begin{figure}[!htbp] \label{Fig1}
    \centering
    \includegraphics[width=0.41\textwidth]{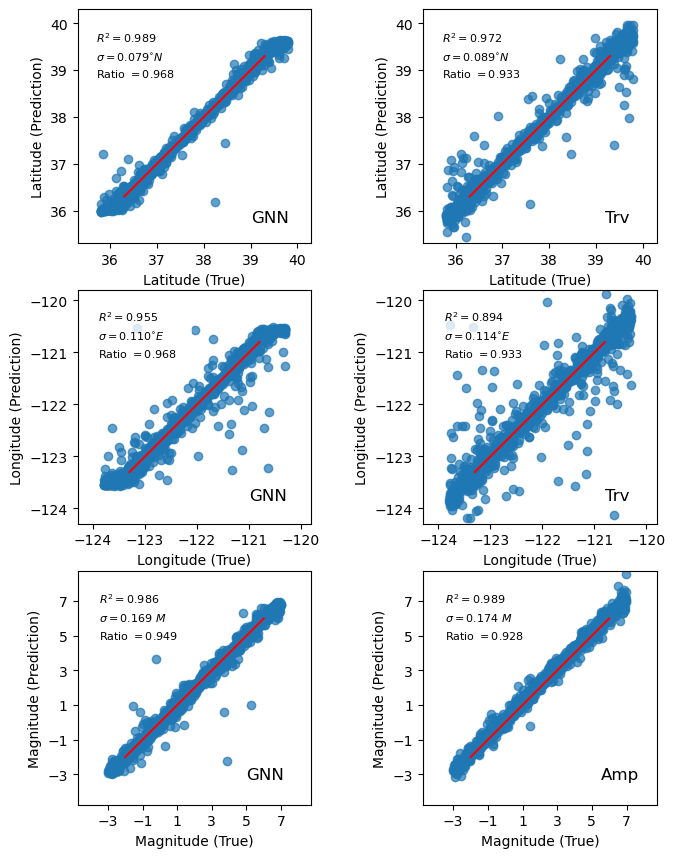}
    \caption{Summary of test results of the trained GNN compared against the traditional location, and magnitude scale methods. In each sub-panel, the $R^2$ correlation scores, standard deviations of residuals of matched events, and the rate of recovered events (locations matched within $\leq 0.5^{\circ}$, and magnitudes matched within $\leq 0.5$ M) are reported.} \vspace{-0.2 cm}
\end{figure}

The results of the location and magnitude analysis are summarized in Fig. 1, where we observe comparable, and slightly favorable performance of the GNN compared with the traditional method in nearly all bulk statistics. Notably, the $R^2$ scores for the GNN are $>$0.95 in latitude and longitude, and improve upon the traditional methods performance by $\sim$0.015, and $\sim$0.05 units, respectively. Also, by declaring `matched events' as those where the predicted event is within $0.5^{\circ}$ epicentral distance of the true location, we measure a recovery rate of 0.968 for the GNN, and 0.933 for the travel time method, highlighting the GNN's improved tolerance to outliers. Both methods also estimate event depths, however because the station spacing is large compared to the range of seismogenic depths ($[0, 40]$ km) and all observations are from the Earth's surface, depths are more poorly constrained by the noisy input data than horizontal coordinates, and have $R^2$ = 0.29 and $R^2$ = 0.22 for the GNN, and travel time method, respectively. For the magnitude application, we find comparable performance between both methods, with the GNN having slightly reduced residuals and rates of missed events. The magnitude residuals of the traditional method are systematically positive by $\sim$0.16 M, while the GNN residuals are well localized around a mean of zero for both magnitude, and location predictions.

\section{Conclusion}

By making use of the local structure in data through information passing along the adjacency matrix, graph-based methods have the potential to parse subtle, but important feature interactions in datasets. Our application demonstrates the potential of GNN's to solve traditional seismological inverse problems, while directly incorporating the physics of the problem both through the defined input features, and the adjacency matrices. The framework we propose can be easily adapted and extended to other similar inverse problems, and the model can also be trained or applied on real data.

\section*{Acknowledgment}

{\small{This work is supported by Department of Energy (Basic Energy Sciences; Award DE-SC0020445).}}

\bibliography{bib}
\bibliographystyle{IEEETrans}

\end{document}